\documentclass[namedreferences]{SolarPhysics}
\usepackage[optionalrh]{spr-sola-addons} 
\usepackage{graphicx}        
\usepackage{color}           
\usepackage{url}             

\ifx \doiurl
 \undefined \def
\doiurl#1{\href{http://dx.doi.org/#1}{\textsf{DOI}}}
\fi

\begin{document}

\begin{article}

\begin{opening}

\title{Growth Rates of the Upper-hybrid Waves for Power-law and Kappa Distributions with a Loss-cone Anisotropy}

\author{L.V. \surname{Yasnov}$^{1,2}$\sep
        J. \surname{Ben\'a\v{c}ek}$^{3}$\sep
        M.   \surname{Karlick\'y}$^{4}$}

        \runningauthor{L.V. Yasnov \textit{et al.}}
        \runningtitle{Growth rates of the Upper-hybrid Waves for Power-law and Kappa ...}

\institute{$^{1}$ St.-Petersburg State
                  University, St.-Petersburg, 198504, Russia
                  \email{l.yasnov@spbu.ru}\\
           $^2$ St.-Petersburg branch of Special Astrophysical Observatory, 196140,  St.-Petersburg,
           Russia\\
           $^{3}$ Department of Theoretical Physics and Astrophysics, Masaryk University,
   Kotl\'a\v{r}sk\'a 2, CZ -- 611 37 Brno, Czech Republic
           \email{jbenacek@physics.muni.cz}\\
           $^{4}$ Astronomical Institute, Academy of Sciences of
           the Czech Republic, 251 65 Ond\v{r}ejov, Czech Republic
           \email{karlicky@asu.cas.cz}\\
             }

\date{Received ; accepted }

%

\begin{abstract}
Fine structures of radio bursts play an important role in diagnostics of the
solar flare plasma. Among them the zebras, which are prevalently assumed to be
generated by the double plasma resonance instability, belong to the most
important. In this paper we compute the growth rate of this instability for two
types of the electron distribution: a) for the power-law distribution and b)
for the kappa distribution, in the both cases with the loss-cone type
anisotropy. We found that the growth rate of the upper-hybrid waves for the
power-law momentum distribution strongly depends on the pitch-angle boundary.
The maximum growth rate was found for the pitch-angle $\theta_\mathrm{c}
\approx$ 50$^\circ$. For small angles the growth rate profile is very flat and
for high pitch-angles the wave absorption occurs.  Furthermore, analyzing the
growth rate of the upper hybrid waves for the kappa momentum distribution we
found that a decrease of the characteristic momentum $p_\kappa$ shifts the
maximum of the growth rate to lower values of the ratio of the electron-plasma
and electron-cyclotron frequencies, and the frequency widths of the growth rate
peaks are very broad.  But, if we consider the kappa distribution which is
isotropic up to some large momentum $p_m$ and anisotropic with loss-cone above
this momentum then distinct peaks of the growth rate appear and thus distinct
zebra stripes can be generated. It means that the restriction for small
momenta for the anisotropic part of distributions is of principal importance
for the zebra stripes generation. Finally, for the 1 August 2010 zebra stripes,
the growth rates in dependence on radio frequency were computed. It was shown
that in this case the growth rate peaks are more distinct than in usually
presented dependencies of growth rates on the ratio of the plasma and cyclotron
frequencies.
\end{abstract}

%
\end{opening}

\section{Introduction}

Zebra structure is a fine structure of Type IV radio bursts observed during
solar flares in the decimetric, metric and centimetric wavelength
ranges
\citep{1972SoPh...25..210S,2012A&A...538A..53C,2012ApJ...744..166T,2014ApJ...780..129T}.
There are many models of this fine structure
\citep{1972SoPh...24..210R,1975SoPh...44..461Z,1975A&A....40..405K,1976SvA....20..449C,1990SoPh..130...75C,2003ApJ...593.1195L,
2006A&A...450..359B,2006SoPh..233..129L,2007SoPh..241..127K,2009PlPhR..35..160L,2010Ap&SS.325..251T,2013A&A...552A..90K},
see also reviews by
\cite{2010RAA....10..821C,2014RAA....14..831C,2016PhyU...59..997Z}. Among these
models the most commonly used model is based on the double-plasma resonance
(DPR) instability, see e.g. the review by \cite{2016PhyU...59..997Z}.

The process of the double plasma resonance, which generates the upper-hybrid
waves, is the most effective in the flare loop regions, where the condition
$\omega_\mathrm{p} \simeq s \omega_\mathrm{B}$ is fulfilled
($\omega_\mathrm{p}$ and $\omega_\mathrm{B}$ means the electron plasma and
electron gyro frequency, $s$ is the gyro-harmonic number). However, this
process strongly depends on distributions of accelerated electrons. In many
papers
\citep{1975SoPh...44..461Z,1986ApJ...307..808W,2004SoPh..219..289Y,2017A&A...555A...1B,2017SoPh..292..163Y,2018A&A...611A..60B}
studying zebra stripes the distribution of accelerated electrons were described
by the Dory-Guest-Harris (DGH) type function \citep{1965PhRvL..14..131D}.
However, this distribution has not a clear physical foundation. The
distributions that are a result of processes of accelerations and reflections
of electrons in magnetic mirrors in closed magnetic loops are physically more
acceptable \citep{1974SvA....17..781S,1974SoPh...36..157K,1983PASAu...5..188W}.
Therefore \cite{1986ApJ...307..808W} considered the loss-cone distribution with
the exponential function of the momentum of electrons. Furthermore, in the
paper by \cite{2007SoPh..241..127K}, the authors considered the loss-cone
distribution with the power-law function of the momentum of electrons. This
distribution is more realistic because it is used in a power-law fitting of
hard X-ray spectra of solar flares. But in this distribution the low-energy
cut-off needs to be defined, which is difficult to estimate from observations
\citep{2003ApJ...595L..97H,2005A&A...435..743S,2008SoPh..252..139K}. Therefore,
in recent years an interest about the kappa distribution is increasing. This
distribution has no low-energy cut-off and is close to Maxwellian
distribution at low energies and at high energies is similar to the power-law
one. Kappa distributions are supported by theoretical considerations of
particle acceleration in collisional plasmas \citep{2014ApJ...796..142B}.
Furthermore, the X-ray spectra of coronal X-ray sources are well fitted using
kappa distributions
\citep{2009A&A...497L..13K,2013ApJ...764....6O,2015ApJ...799..129O}.

In the present article, firstly, we follow the study of
\cite{2007SoPh..241..127K}. We extend their analysis in order to show changes
of the growth rate for the power-law momentum distribution in dependence on the
low-energy cut-off and loss-cone angle. Then we present growth rates for the
anisotropic kappa distribution and kappa distribution which is isotropic up to
some large momentum and anisotropic above this momentum. Finally, for the first
time, for the zebra stripes observed at 1 August 2010, we compute the growth
rates in dependence on radio frequency.

\section{Growth Rates for Power-law Distributions}

\begin{figure}
\centering
\includegraphics[width=10cm]{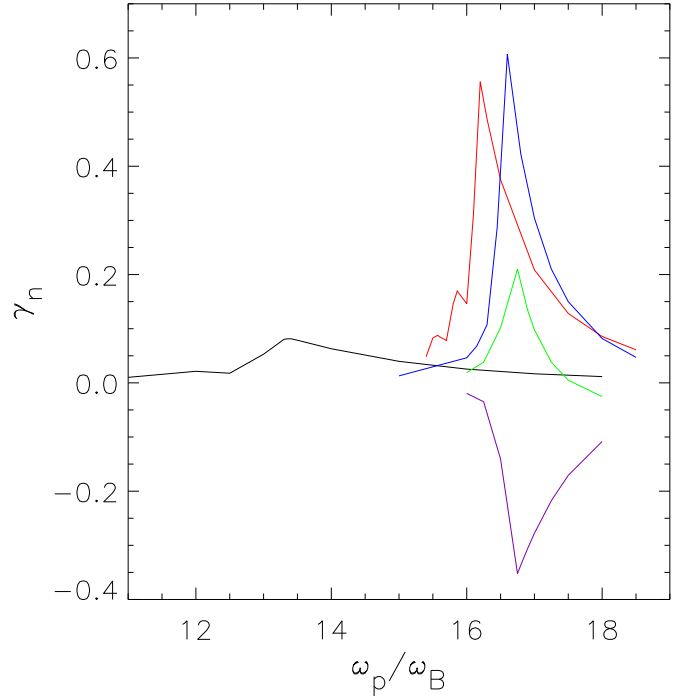}
\caption{Maximum growth rate of the upper-hybrid waves for the power-law momentum distribution
in dependence on the ratio $\omega_\mathrm{p}/\omega_\mathrm{B}$ and the loss-cone angle 10$^\circ$ (black line), 30$^\circ$ (red line),
50$^\circ$ (blue line), 65$^\circ$ (green line) and 80$^\circ$ (violet line).
The power-law index of the power-law distribution is $\delta$ = 5, the gyro-harmonic number is $s$ =16 and
the minimum electron momentum p$_\mathrm{m}$ corresponds to the velocity 0.3 c, i.e. to the low-energy cut-off $\approx$ 30 keV.}
\label{figure1}
\end{figure}

\begin{figure}
\centering
\includegraphics[width=6cm]{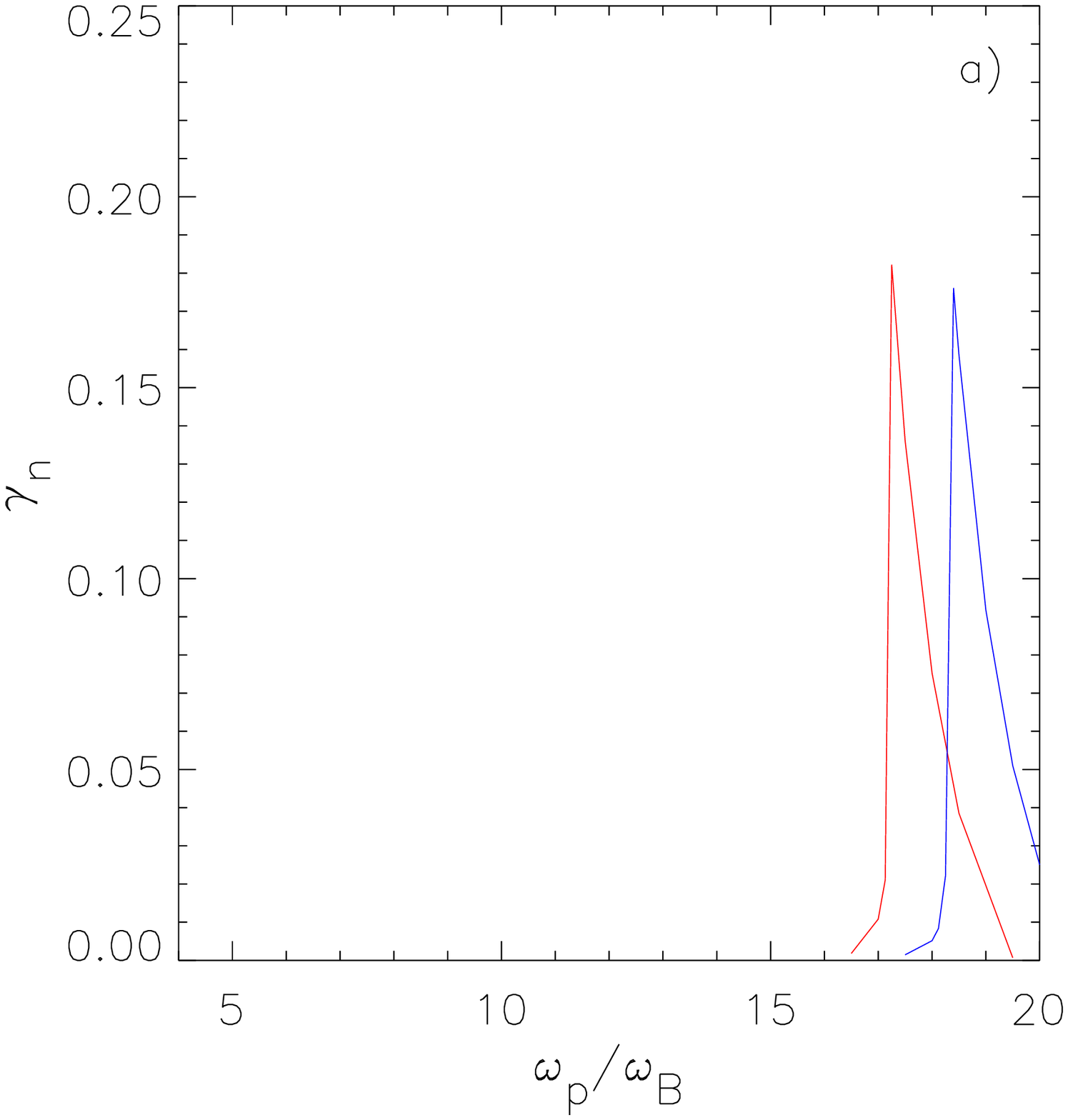}
\includegraphics[width=6cm]{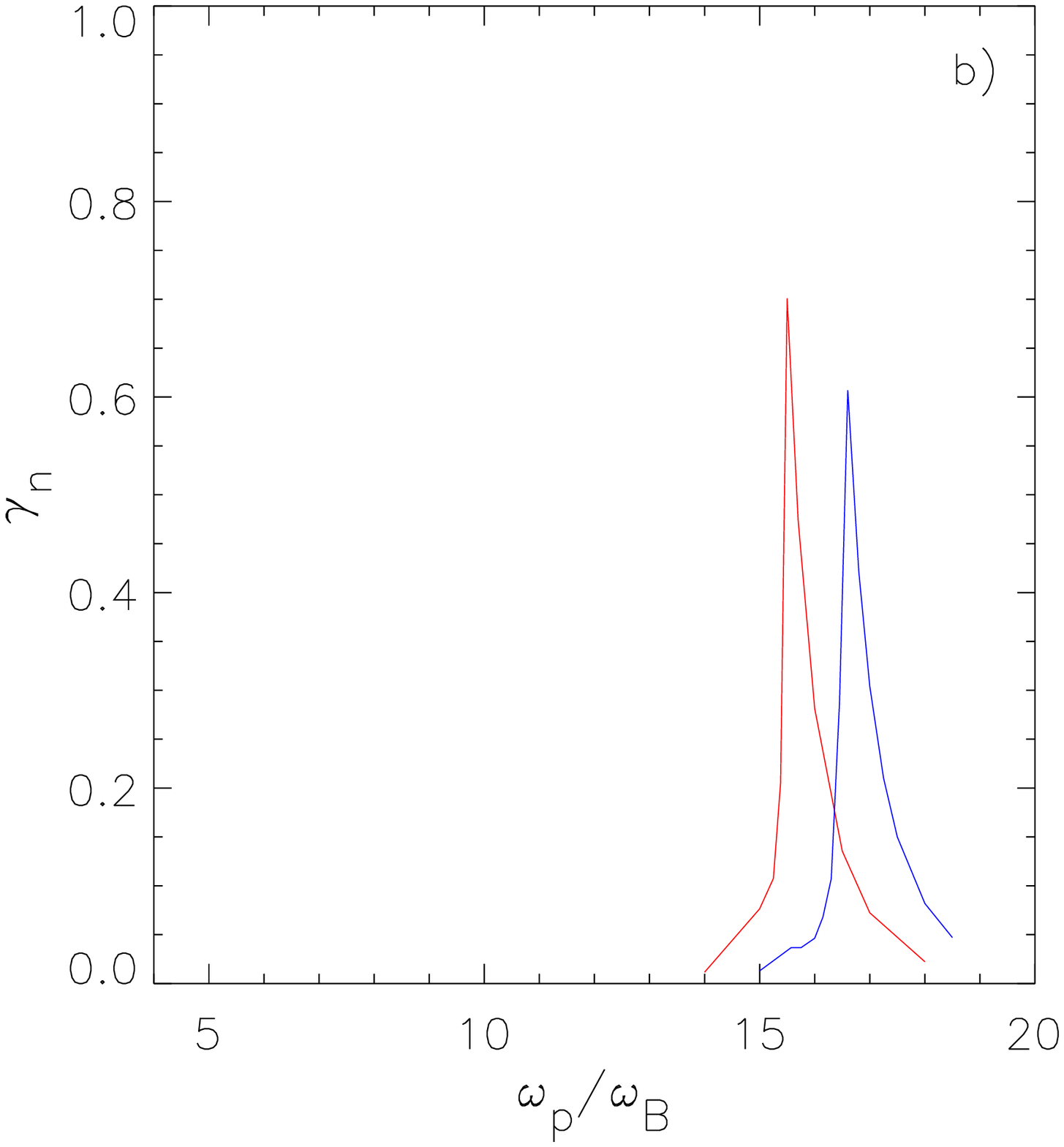}
\includegraphics[width=6cm]{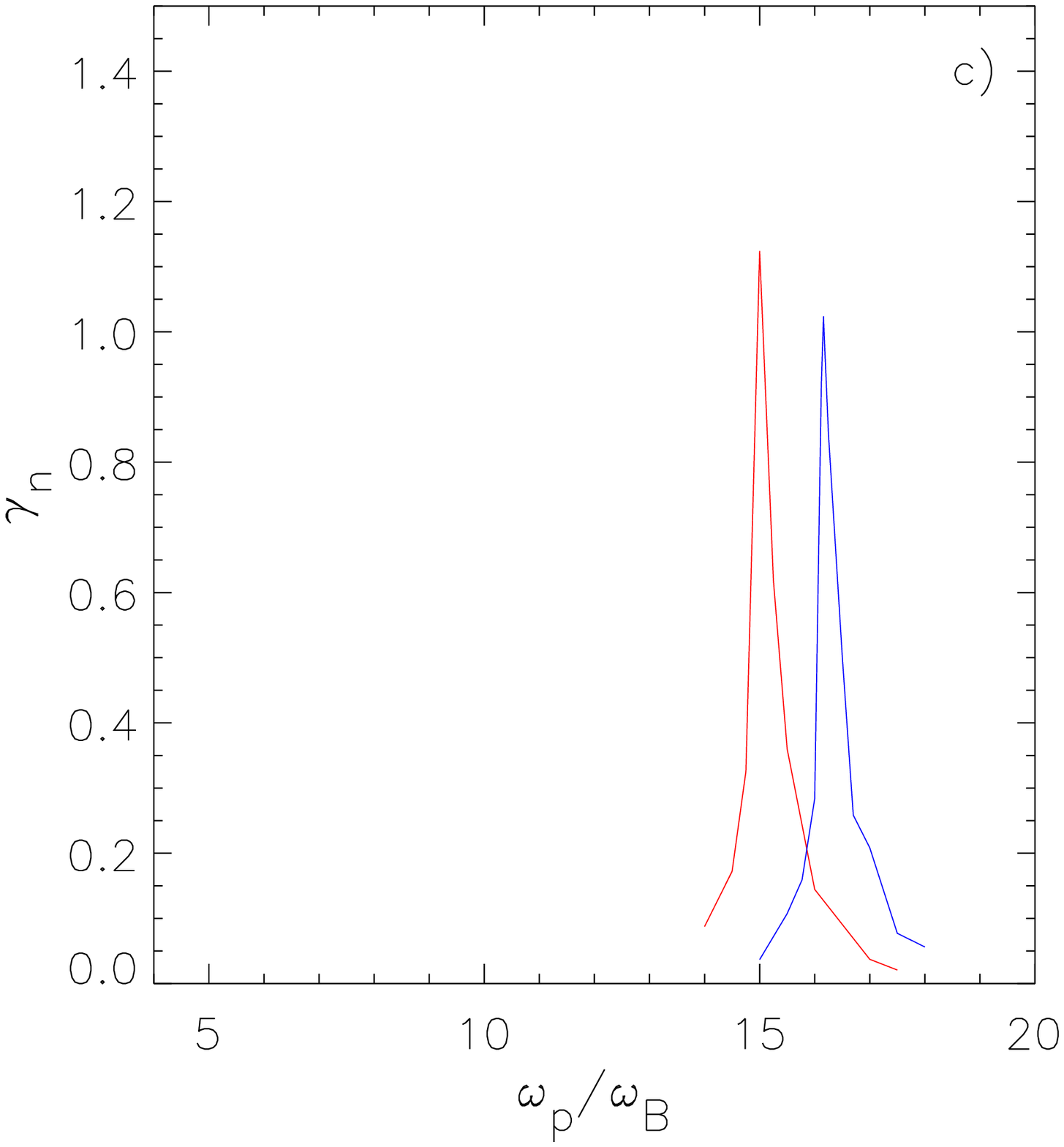}
\includegraphics[width=6cm]{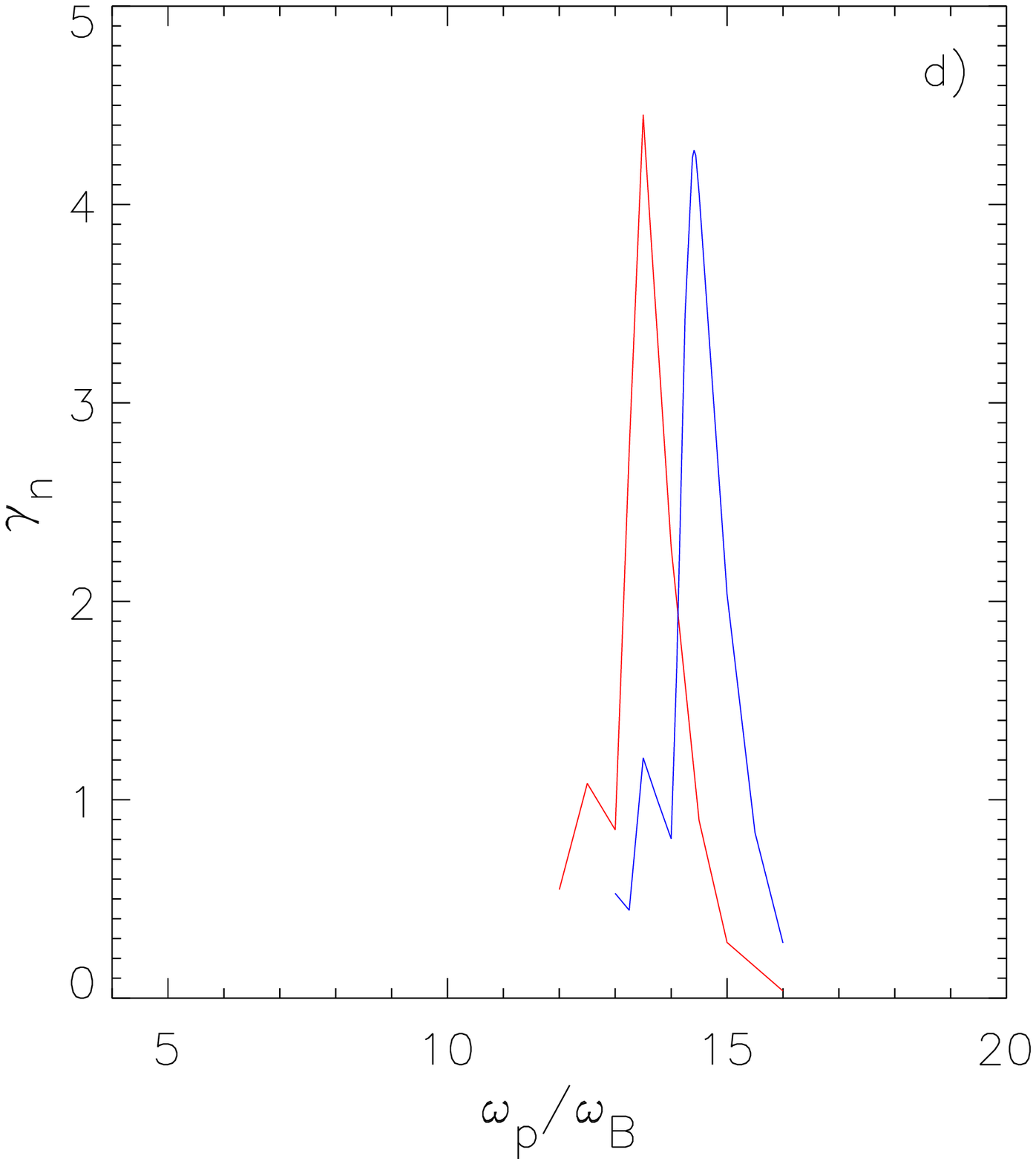}
\includegraphics[width=6cm]{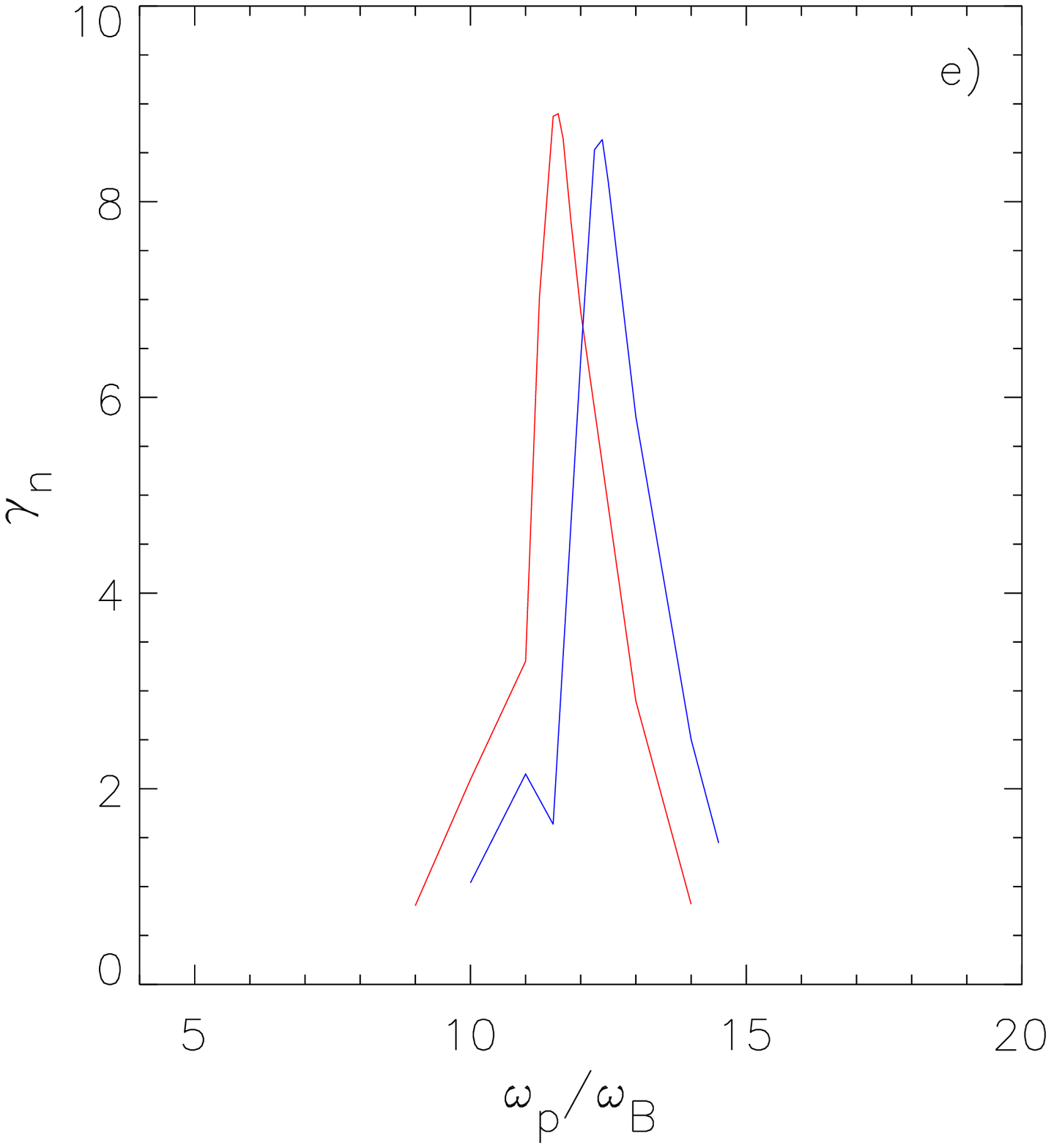}
\includegraphics[width=6cm]{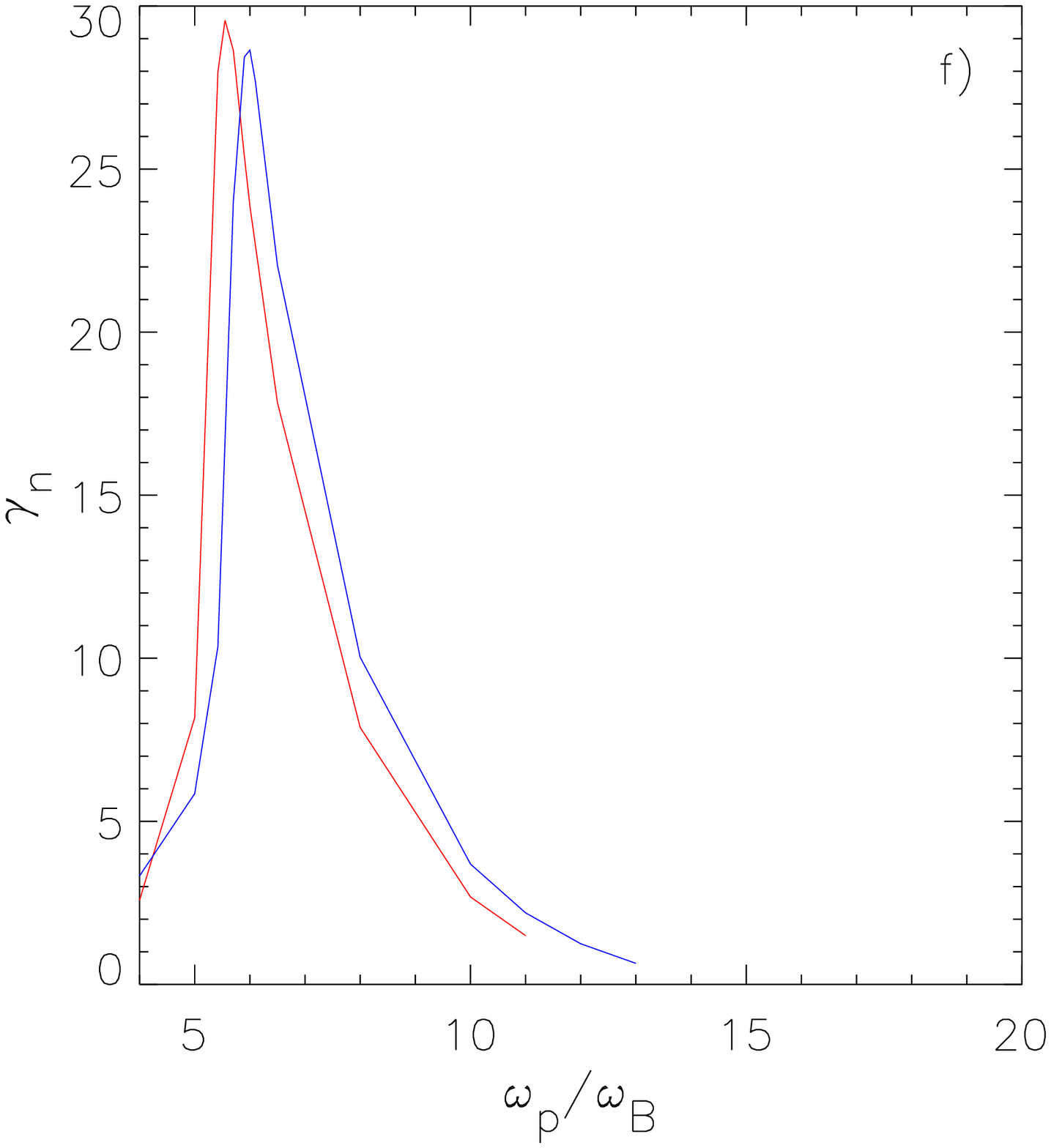}
\caption{Maximum growth rate of the upper-hybrid waves for the power-law momentum distribution
in dependence on the ratio $\omega_\mathrm{p}/\omega_\mathrm{B}$ and the minimum electron momentum p$_\mathrm{m}$
for the gyro-harmonic number $s$ = 15 (red line) and $s=16$ (blue line).
Plot a) is for p$_\mathrm{m}$ corresponding to the velocity
0.5 c, b) for p$_\mathrm{m}$ corresponding to 0.3 c,
c) for p$_\mathrm{m}$ corresponding to 10 v$_\mathrm{T}$, d) for p$_\mathrm{m}$ corresponding to 5 v$_\mathrm{T}$,
e) for p$_\mathrm{m}$ corresponding to 3 v$_\mathrm{T}$,
and f) for p$_\mathrm{m}$ corresponding v$_\mathrm{T}$. The power-law index is $\delta$ = 5 and
the loss-cone is $\theta_\mathrm{c}$ = 50$^\circ$. Note that the scale on the y-axis increases from plot a) to f).}
\label{figure2}
\end{figure}

\begin{figure}
\centering
\includegraphics[width=10cm]{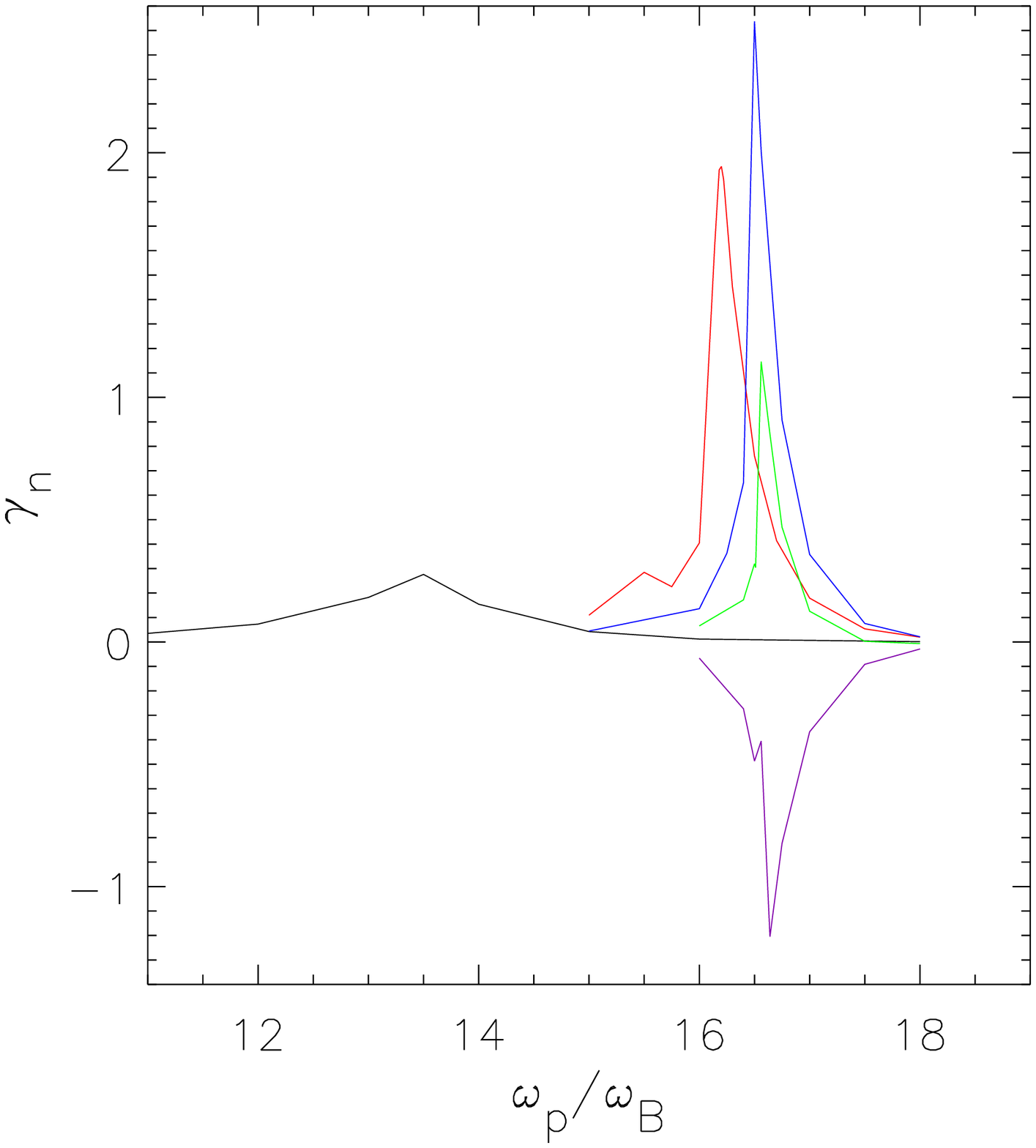}
\caption{Maximum growth rate of the upper-hybrid waves for the power-law momentum distribution
in dependence on the ratio $\omega_\mathrm{p}/\omega_\mathrm{B}$ and the loss-cone angle 10$^\circ$ (black line), 30$^\circ$ (red line),
50$^\circ$ (blue line), 65$^\circ$ (green line) and 80$^\circ$ (violet line).
The power-law index of the power-law distribution is $\delta$ = 10, the gyro-harmonic number is $s$ =16 and
the minimum electron momentum p$_\mathrm{m}$ corresponds to the velocity 0.3 c, i.e. to the low-energy cut-off $\approx$ 30 keV.}
\label{figure2b}
\end{figure}

Let us briefly describe the growth rate calculation. We follow the method
according to \cite{2007SoPh..241..127K}. We consider a plasma with two
components: a) the background Maxwellian plasma and b) hot non-equilibrium
plasma component with the plasma densities $n_\mathrm{0}$ and $n_\mathrm{h}$, respectively, where
$n_\mathrm{0} \gg n_\mathrm{h}$. The electron distribution function of this hot non-equilibrium
component is taken as
\begin{equation}
f(p, \theta) = \varphi(p)\left\{ \begin{array}{lll}
0, & \theta \le \theta_\mathrm{c} - \Delta\theta_\mathrm{c}, \\
\frac{\theta - \theta_\mathrm{c} + \Delta \theta_\mathrm{c}}{\Delta \theta_\mathrm{c}}, & \theta_\mathrm{c} - \Delta \theta_\mathrm{c} < \theta < \theta_\mathrm{c}, \\
        1, & \theta> \theta_\mathrm{c}.
        \end{array}
\right.
\label{ee1}
\end{equation}

This function describes the distribution with the loss-cone having the pitch
angle boundary $\theta_\mathrm{c}$ and the boundary width $\Delta
\theta_\mathrm{c} \ll$ 1. The function $\varphi(p)$ describes the distribution
in dependence on the electron momentum.

Here the $\varphi(p)$ function is taken in the power-law function form:
\begin{equation}
\varphi(p) = \left\{ \begin{array}{ll}
\frac{\delta - 3}{2 \pi (\pi - \theta_\mathrm{c})p_\mathrm{m}^3} \left( \frac{p}{p_\mathrm{m}} \right)^{-\delta} & p \geq p_\mathrm{m} \\
0  & p < p_\mathrm{m},
\end{array}
\right.
\end{equation}
where $p_\mathrm{m}$ is the low-momentum cut-off, and $\delta$ is the power-law index.
Note that this distribution is normalized to one.

Generally, for the growth rate of the upper-hybrid waves we can write
\begin{equation}
\gamma = -\frac{\mathrm{Im\, \epsilon_\parallel}}{ \left. \frac{\partial \mathrm{Re \epsilon_{\parallel}}}{\partial \omega} \right\vert_{\epsilon_\parallel = 0} },
\end{equation}
\begin{equation}
\left. {\frac{\partial \mathrm{Re}\, \epsilon_\parallel}{\partial \omega}}\right\vert_{\mathrm{\epsilon_\parallel = 0}} \simeq \frac{2}{\omega} \left( 2 - \frac{\omega_\mathrm{p}^2}{\omega^2} \right),
\end{equation}
where $\epsilon_\parallel$ is dielectric permeability and $\omega$ is the frequency of the upper-hybrid
waves.

For the term $\mathrm{Im\, \epsilon_\parallel}$ we use the relation (17) from the
paper by \cite{2007SoPh..241..127K}
\begin{eqnarray}
\mathrm{Im}\, \epsilon_\parallel^{(s)} \simeq -2\pi^2 m^4 c^2 \frac{\omega_\mathrm{p}^2}{k^2}\frac{n_\mathrm{h}}{n_\mathrm{0}}
 \Gamma_\mathrm{r}^3 J_s^2 \left( \frac{k_\perp p_\perp}{m\, \omega_\mathrm{B}} \right) \times \nonumber \\
\times \left[ \frac{\partial \varphi(p)}{\partial p}  + \frac{\varphi(p) \tan \theta_\mathrm{c}}
{p \Delta \theta_\mathrm{c}} \left( \frac{s \omega_\mathrm{B}}{\Gamma_\mathrm{r} \omega \sin^2 \theta_\mathrm{c}} - 1 \right) \right]
 \frac{\Delta p_\mathrm{z}}{p_\mathrm{0}},
\end{eqnarray}
where $J_s$ is the s-th order Bessel function, $\omega_\mathrm{B}$ is the
electron cyclotron frequency, $\mathbf{p} = (p_\perp, p_{z0}) = (p_\mathrm{0}
\sin \theta_\mathrm{c}, p_\mathrm{0} \cos \theta_\mathrm{c})$ is the electron
momentum, $m$ is the electron mass, $c$ is the speed of light, $k$ is the wave
number. The distance in momentum space $\Delta p_\mathrm{z}$ between
intersection points with straight line for small parameter $\Delta
\theta_\mathrm{c} \ll 1$ is
\begin{equation}
\Delta p_\mathrm{z}= 2 p_\mathrm{z0} \frac{\omega}{s \omega_\mathrm{B}} \sqrt{2 \Delta \theta_\mathrm{c} \tan \theta_\mathrm{c} }.
\end{equation}

Then the normalized growth rate can be expressed in agreement with the paper by
\cite{2007SoPh..241..127K} as
\begin{eqnarray}
 \gamma_\mathrm{n} = \frac{\gamma}{\omega_\mathrm{p}} \frac{n_\mathrm{0}}{n_\mathrm{h}} \sqrt{\Delta \theta_\mathrm{c}} = \frac{4}{\sqrt{2}}
 \pi^2 m^4 c^2 \frac{\omega\, \omega_\mathrm{p}}{ k^2} \Gamma_\mathrm{r}^4 J_s^2
 \left( \frac{k_\perp p_\mathrm{0} \sin \theta_\mathrm{c}}{m \omega_\mathrm{B}} \right) \times \nonumber \\
 \times \varphi(p_\mathrm{0}) \frac{ \tan^{3/2}(\theta_\mathrm{c}) \cos(\theta_\mathrm{c}) }
 {p_\mathrm{0} (2 - \frac{\omega_\mathrm{p}^2}{\omega^2})} \left( \frac{1}{\Gamma_\mathrm{r}^2 \sin^2 \theta_\mathrm{c}} - 1\right),
\label{eq1}
\end{eqnarray}
where $p_\mathrm{0}$ is the lower boundary for hot electron momentum
\begin{equation}
p_\mathrm{0} = \frac{m c \sqrt{\omega^2 - s^2 \omega_\mathrm{B}^2}}{s \omega_\mathrm{B}},
\end{equation}
$\mathbf{k} = (k_\mathrm{z}, k_\perp)$ is the
  wave vector with the components along and in the perpendicular direction to magnetic field
\begin{eqnarray}
\mathbf{k}^2 &=& k_\mathrm{z}^2 + k_{\perp}^2 = \frac{ \omega^2  - s^2 \omega_\mathrm{B}^2}{c^2 \cos^2(\theta_\mathrm{c})} + \frac{\omega^4 - \omega^2 \omega_\mathrm{p}^2 - \omega_\mathrm{B}^2 \omega_\mathrm{p}^2}{3 v_\mathrm{T}^2 \omega_\mathrm{p}^2 },
\end{eqnarray}
$\Gamma_\mathrm{r}$ is the relativistic factor
\begin{equation}
\Gamma_\mathrm{r} = \frac{\omega }{s \omega_\mathrm{B}},
\end{equation}
and $v_\mathrm{T}$ is the thermal velocity of the background plasma.

Now using the relation (\ref{eq1}) we computed the growth rate for the
following parameters: The lower limit for the momentum of electrons $p_\mathrm{m}$ is
taken as corresponding to the minimum energy $E_\mathrm{m}$ $\approx$ 30 keV, i.e., to
the minimum velocity of electrons $v_\mathrm{m}/c$ = 0.3, the power-law index is
$\delta$ = 5 and the gyro-harmonic number is $s$ = 16. The pitch-angle boundary
varies as $\theta_\mathrm{c}$ = 10$^\circ$, 30$^\circ$, 50$^\circ$, 65$^\circ$ and
80$^\circ$ and the temperature of the background plasma is $T_\mathrm{0}$ = 3 $\times$
10$^6$ K.

In computations we only varied the magnetic field $B$, while the plasma
frequency was kept constant ($f_\mathrm{p} = \omega_\mathrm{p}/2 \pi$ = 1 GHz).
For each value of $\omega_\mathrm{p}/\omega_\mathrm{B}$ the growth rate
$\gamma_\mathrm{n}$ was computed in the frequency interval
$\sqrt{\omega_\mathrm{p}^2 + \omega_\mathrm{B}^2} < \omega \leq
\omega_\mathrm{max}$. The frequency $\omega_\mathrm{max}$ was taken by an
experimental way in order to find the maximum value of $\gamma_\mathrm{n}$ in
this interval. The results of these computations are shown in
Figure~\ref{figure1}. As seen here the growth rate strongly depends on the
value of the pitch-angle boundary. The maximum peak is for $\theta_\mathrm{c}
\approx$ 50$^\circ$. For small angles no distinct peak is visible, and for high
angles the absorption appears.

Now, let us analyze an effect of variation of $p_\mathrm{m}$ (i.e., the
low-velocity limit of electrons) on the growth rate. The result for $\delta$ =
5 and $\theta_\mathrm{c}$ = 50$^\circ$ is shown in Figure~\ref{figure2}. The
value of $p_\mathrm{m}$ varies in correspondence with the minimum electron
velocity $v_\mathrm{m} \in$ (0.5 c -- $v_\mathrm{T}$), $v_\mathrm{T}$ = 6.75
$\times$ 10$^6$ m s$^{-1}$. Figure~\ref{figure2} shows that the maximum
contrast between peaks is for the growth rates with the minimum electron
velocities in the 0.3 c -- 10 $v_\mathrm{T}$ range, see also Table~\ref{tab1}.
For velocities greater than 0.5 c the contrast of peaks decreases and for the
velocities $\leq$ 5 $v_\mathrm{T}$ the peaks are shifted to much lower ratio of
$\omega_\mathrm{p}/\omega_\mathrm{B}$. If we accept that the growth rate
profiles correspond to the intensity of zebra stripes, it means that for the
low $p_\mathrm{m}$ no zebra stripes can be generated.

The same computations were made also for the power-law distribution with the
power-law index $\delta$ = 10. The computed growth rates for this power-law
distribution with different pitch angles are shown in Figure~\ref{figure2b}.
Similarly as in the case with the power-law distribution with the power-law
index $\delta$ = 5, we can see a strong dependence on the pitch angle, however,
the maximum is one again for the pitch angle $\theta_\mathrm{c} \approx$
50$^\circ$. On the other hand, the growth rates are higher and narrower in
frequency comparing with the previous case.

Summarizing all these results, in Table~\ref{tab1} we present the ratio
$(\omega_\mathrm{p}/\omega_\mathrm{B})_\mathrm{m}$, where the growth rate has
maximum and the peak width $\Delta(\omega_\mathrm{p}/\omega_\mathrm{B})$ (taken
at half of the maximum) in dependence on the minimum electron velocity
$v_\mathrm{m}$ (minimum of $p_\mathrm{m}$) for the gyro-harmonic number $s$ =
16 and for the power-law index $\delta$ = 5 and 10. To see separate zebra
stripes the peak width needs to be smaller than 0.5. While for the power-law
index $\delta$ = 5 the zebra structure can be formed only in limited interval
of $v_\mathrm{m}$ around the velocity $10\,v_\mathrm{T}$ = 6.75 $\times$ 10$^7$
m s$^{-1}$, in the case with $\delta$ = 10 the width of the growth rates are
two time smaller and thus more favorable for the zebra pattern generation.
Positions of the growth rate maxima in both the cases are approximately the
same.

\begin{table}
\caption{Frequency ratio of the growth rate maximum
$(\omega_\mathrm{p}/\omega_\mathrm{B})_\mathrm{m}$ and the bandwidth of the
growth rate peak $\Delta(\omega_\mathrm{p}/\omega_\mathrm{B})$ for the harmonic
number $s=16$ in dependence on the minimum velocity $v_\mathrm{m}$
(corresponding to $p_\mathrm{m}$) of the power-law distribution for two
power-law indices ($\delta$ = 5 and 10).} \centering
\begin{tabular}{ccccccc}
\hline
\hline
 $\delta$ & $v_\mathrm{m}$ & 0.5 c  & 0.3 c & 10 v$_\mathrm{T}$ & 5 v$_\mathrm{T}$ & v$_\mathrm{T}$  \\
\hline
5 & $(\omega_\mathrm{p}/\omega_\mathrm{B})_\mathrm{m}$     & 18.41 & 16.60 & 16.16 & 14.42 & 5.98 \\
5 & $\Delta(\omega_\mathrm{p}/\omega_\mathrm{B})$ & 0.73  & 0.56  & 0.46  & 0.84  & 1.90  \\
\hline
10  & $(\omega_\mathrm{p}/\omega_\mathrm{B})_\mathrm{m}$     & 18.43 & 16.50 & 15.95 & 14.41 & 6.00 \\
10  & $\Delta(\omega_\mathrm{p}/\omega_\mathrm{B})$ & 0.40  & 0.26  & 0.25  & 0.61  & 1.19  \\
\hline \label{tab1}
\end{tabular}
\end{table}

\section{Growth Rates for Kappa Distributions}

Now, we consider a plasma with the hot component having the kappa distribution
with the loss-cone anisotropy for all electron momentums. The kappa
distribution is taken as \citep{2014ApJ...796..142B}

\begin{equation}
f_\kappa (v) = \frac{n_{\kappa} \Gamma(\kappa + 1)}{\pi^{\frac{3}{2}} \theta_{\kappa}^3 \kappa^{\frac{3}{2}} \Gamma(\kappa - \frac{1}{2})} \left( 1 + \frac{v^2}{\kappa \theta_{\kappa}^2} \right)^{-\kappa -1},
\label{eq11}
\end{equation}
where $\kappa$ is the kappa index, $n_{\kappa} = \int f d^3 v$,
\begin{equation}
\theta_\kappa^2 = \frac{2 k_\mathrm{B} T_\kappa}{m} \frac{\kappa - \frac{3}{2}}{\kappa},
\end{equation}
is the characteristic velocity, $m$ is the electron mass, $k_\mathrm{B}$ is the
Boltzmann constant, $T_\kappa$ is the mean kinetic temperature and $\Gamma$ is
Gamma function.

Similarly as in the calculation of the growth rate for the power-law
distribution we assume the loss-cone type distribution according to the
relation~\ref{ee1}. However in this case the function $\varphi (p)$, derived
from the relation~\ref{eq11}, has the form
\begin{equation}
\varphi(p) = \frac{2 \Gamma(\kappa + 1)}{\pi^\frac{3}{2} p_\kappa^3 \kappa^\frac{3}{2} \Gamma(\kappa - \frac{1}{2})(\pi - \theta_\kappa)} \left( 1 + \frac{p^2}{\kappa p_\kappa^2} \right)^{-\kappa - 1}.
\end{equation}
where $p$ is the electron momentum and $p_\kappa$ = $m \theta_\kappa$.

\begin{figure}
\centering
\includegraphics[width=8cm]{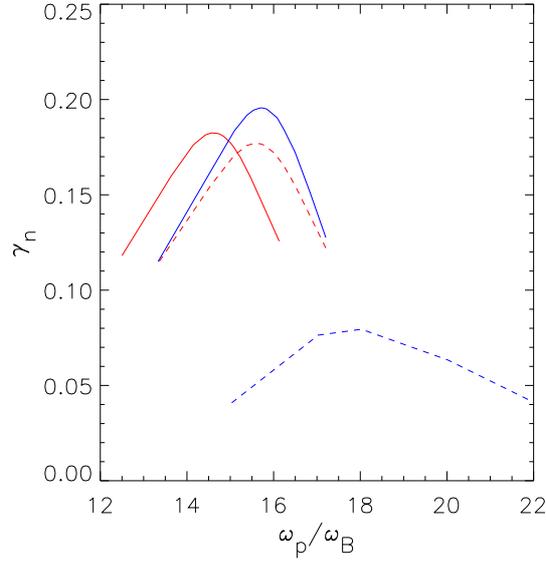}
\caption{Maximum growth rate of the upper-hybrid waves for the kappa momentum distribution
in dependence on the ratio $\omega_\mathrm{p}/\omega_\mathrm{B}$ for the kappa index $\kappa$ = 1.5 and the gyro-harmonic number $s$ = 15 (red solid line),
$k$ = 1.5 and $s$ = 16 (red dotted line), $k$ = 4 and $s$ = 15 (blue solid line),
and $\kappa$ = 4 and $s$ = 16 (blue dotted line).
The loss-cone is $\theta_\mathrm{c}$ = 30$^\circ$, and $p_\kappa$ corresponds to the velocity
0.3 c.} \label{figure3}
\end{figure}

\begin{figure}
\centering
\includegraphics[width=6cm]{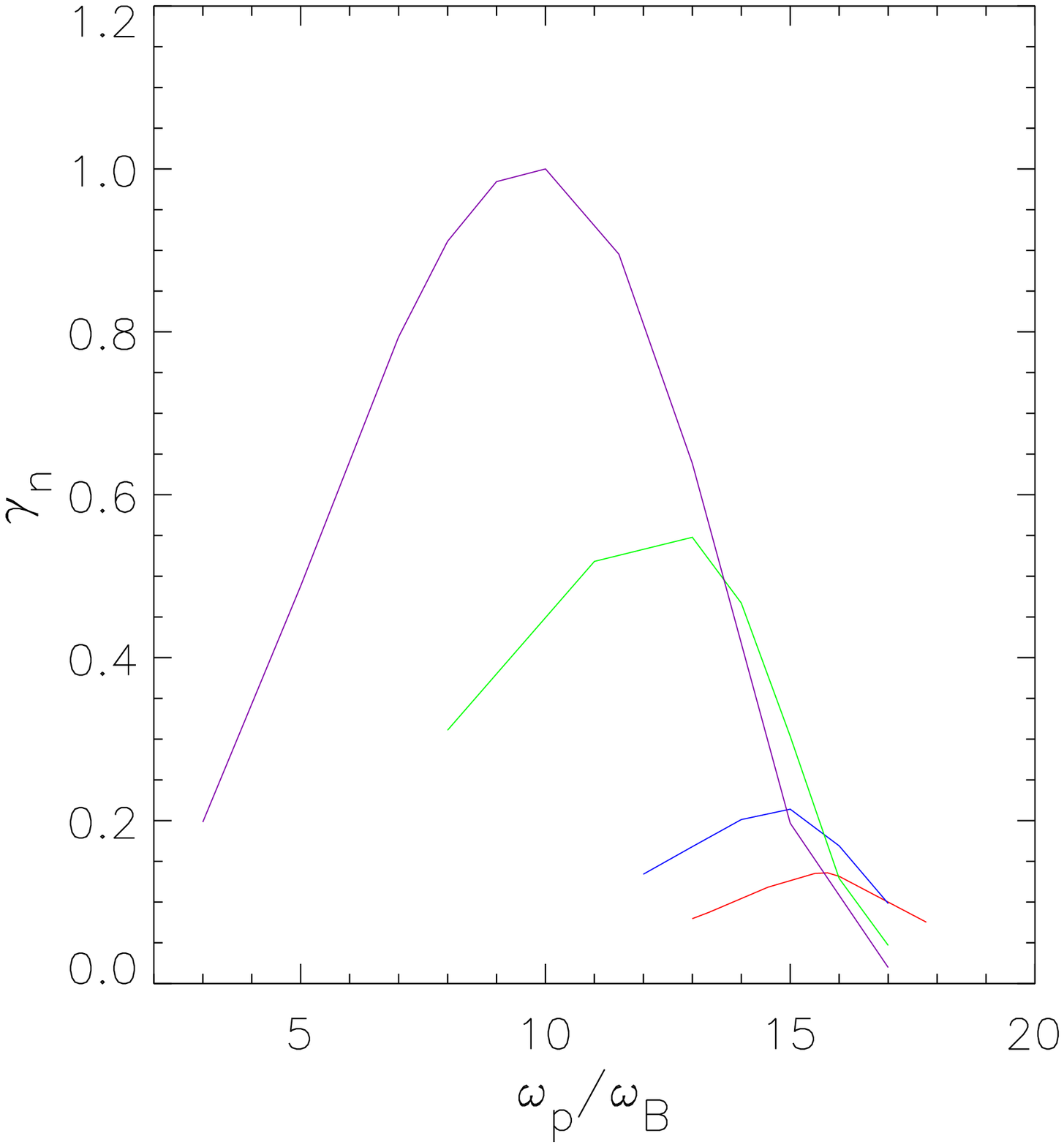}
\includegraphics[width=6cm]{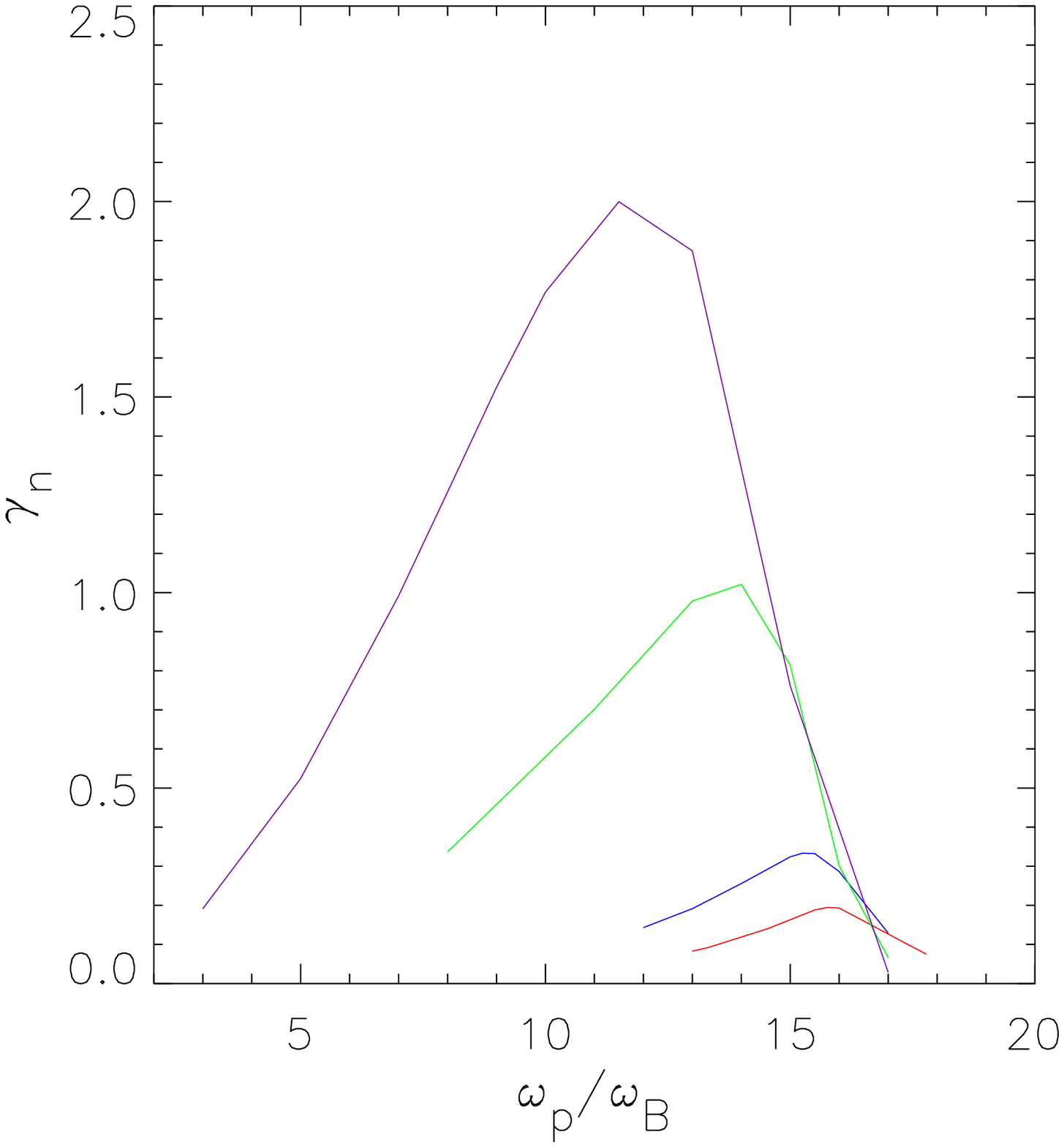}
\caption{Left: Maximum growth rate of the upper-hybrid waves for the kappa momentum distribution
in dependence on the $\omega_\mathrm{p}/\omega_\mathrm{B}$. The kappa index is $\kappa$ = 1.5, the gyro-harmonic number is $s$ = 16 and
the loss-cone is $\theta_\mathrm{c}$ = 30$^\circ$. The red line is for $p_\kappa$ corresponding to 0.3 c,
blue line for p$_\kappa$ corresponding to 10 v$_\mathrm{T}$, green line for $p_\kappa$ corresponding to 5 v$_\mathrm{T}$,
and violet line for p$_\kappa$ corresponding to 3 v$_\mathrm{T}$. Right: The same, but for $\theta_\mathrm{c}$ = 50$^\circ$.}
\label{figure4}
\end{figure}

\begin{figure}
\centering
\includegraphics[width=8cm]{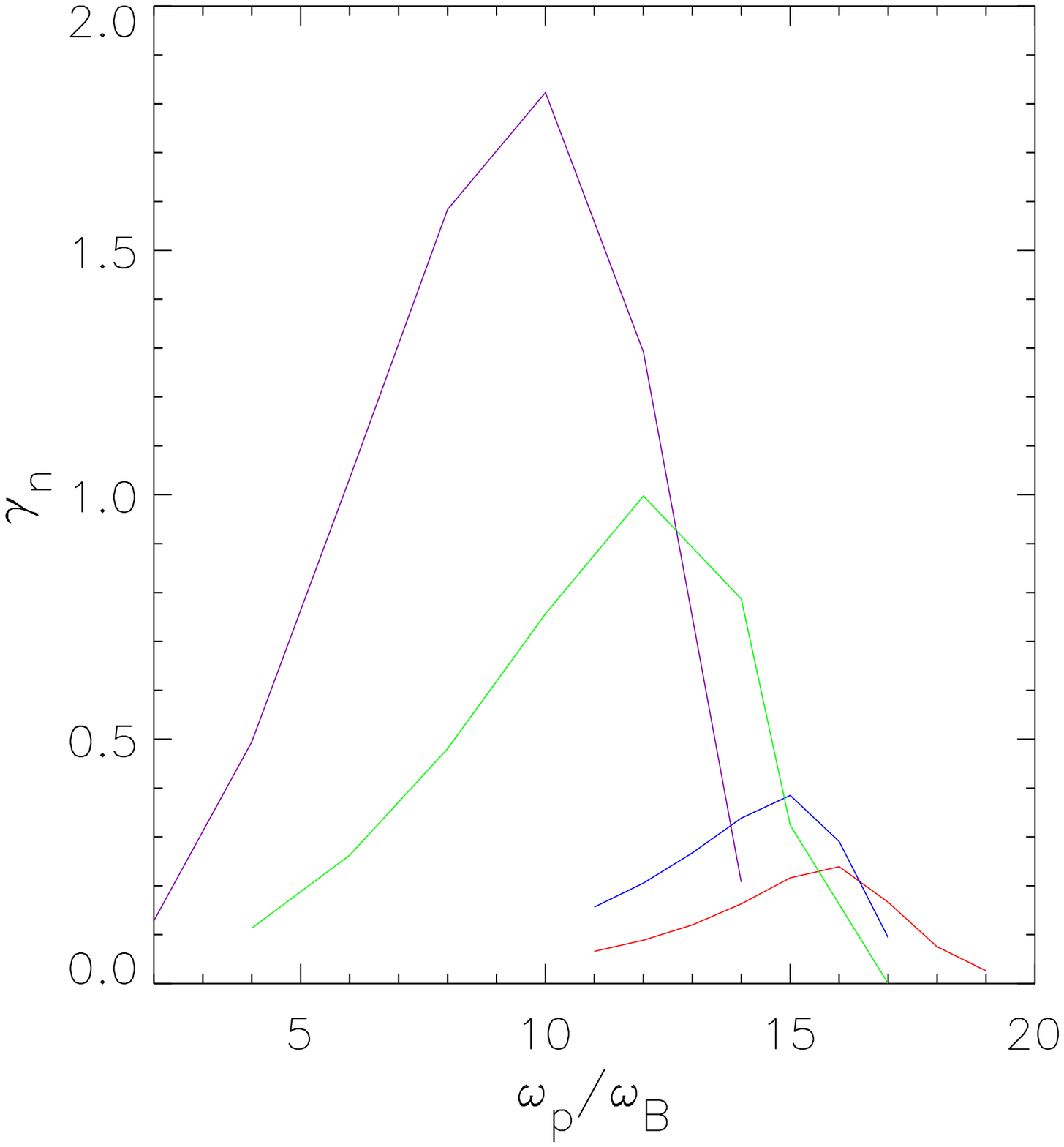}
\caption{Maximum growth rate of the upper-hybrid waves for the kappa momentum distribution
in dependence on the $\omega_\mathrm{p}/\omega_\mathrm{B}$. The kappa index is $\kappa$ = $\infty$, the gyro-harmonic number is $s$ = 16 and
the loss-cone is $\theta_\mathrm{c}$ = 30$^\circ$. The red line for $p_\kappa$ corresponding to 0.3 c,
blue line for $p_\kappa$ corresponding to 10 v$_\mathrm{T}$, green line for $p_\kappa$ corresponding to 5 v$_\mathrm{T}$,
and violet line for $p_\kappa$ corresponding to 3 v$_\mathrm{T}$.}
\label{figure5}
\end{figure}

Using these relations we computed the maximum growth rates for the kappa
momentum distributions with the loss-cone anisotropy. The results are shown in
Figures~\ref{figure3},~\ref{figure4} and~\ref{figure5}. Figure~\ref{figure3}
presents the growth rate in dependence on the ratio
$\omega_\mathrm{p}/\omega_\mathrm{B}$ for the kappa distribution with the kappa
index $\kappa$ = 1.5 (it corresponds to $\delta$ = 5 for the power law
distribution) for the gyro-harmonic numbers $s$ = 15 (red solid line) and 16
(red dashed line), $\theta_\mathrm{c}$ = 30$^\circ$, and $p_\kappa$
corresponding to the velocity 0.3 c). The growth rate for the same parameters,
but for $\kappa$ = 4 are expressed by the blue solid line for $s$ = 15 and by
the blue dashed line for $s$ = 16. As seen in this figure when the kappa index
increases, i.e. the kappa distribution becomes more closer to Maxwellian one,
the bandwidth of the growth rate peaks are broader. Furthermore, while the
values of growth rates for $s$ = 15 and 16 and $\kappa$ = 1.5 are similar, the
value of the growth rate for $s$ = 16 and $\kappa$ = 4 is much smaller
comparing to that with $s$ = 15.

Furthermore, in Figure~\ref{figure4} we show the dependence of the growth rate
for the kappa distribution in dependence on the ratio
$\omega_\mathrm{p}/\omega_\mathrm{B}$ and  $p_\kappa$ for two values of the
loss-cone angle $\theta_\mathrm{c} = 30^\circ$ and $50^\circ$. As seen in
both these figures when we decrease $p_\kappa$ the maximum of the growth rate
increases and shifts to lower values of the ratio
$\omega_\mathrm{p}/\omega_\mathrm{B}$. The growth rates for $\theta_\mathrm{c}
= 50^\circ$ are greater than those for $\theta_\mathrm{c} = 30^\circ$.

Finally, for comparison with Figure~\ref{figure4} left, in Figure~\ref{figure5}
we added the growth rates for the kappa distribution in dependence on the ratio
$\omega_\mathrm{p}/\omega_\mathrm{B}$ for the kappa index $\kappa$ = $\infty$.
The plots of the growth rates are similar, but the values of the growth rates
for $\kappa = \infty$ are higher.

In both Figures~\ref{figure4} and ~\ref{figure5} in all cases the peaks of the
growth rates are very broad. If we assume that the growth rate profiles
correspond to radio emission, it means that in the case with the kappa momentum
distribution distinct zebra stripes cannot be generated.

Therefore now we consider more realistic case, namely, the kappa distribution
in all momentums ($p>0$), but isotropic up to some large momentum $p_\mathrm{m}$ and
anisotropic above this momentum $p_\mathrm{m}$. Now the isotropic part of the kappa
distribution plays a role of the dense background plasma and the anisotropic
part with the kappa momentum distribution and loss-cone anisotropy plays a role
of the low density hot component. Such a division of the distribution is
possible due to the fact that only the anisotropic part of this distribution is
important for the growth rate of the upper-hybrid waves. The isotropic part of
the kappa distribution does not contribute to the growth rate. Therefore for
the following computations only the anisotropic part with momentums above $p_\mathrm{m}$
needs to be expressed. For it we take the distribution, which is normalized to
1, as follows.
\begin{equation}
\varphi(p) = \left\{ \begin{array}{ll}
\frac{(2 \kappa-1) \left(\frac{p^2}{p_\mathrm{m}^2}+\frac{\kappa p_\kappa^2}{p_\mathrm{m}^2} \right)^{-\kappa-1}}{2 \pi  (\pi
    -\theta_\mathrm{c}) p_\mathrm{m}^3 \, _2F_1\left(\kappa-\frac{1}{2};\kappa+1;\kappa+\frac{1}{2};-\frac{\kappa p_\kappa^2}{p_\mathrm{m}^2} \right)}, & p > p_\mathrm{m}, \\
0, & p \le p_\mathrm{m},
\end{array}
\right.
\label{ee2}
\end{equation}
where
$_2F_1\left(\kappa-\frac{1}{2};\kappa+1;\kappa+\frac{1}{2};-\kappa\right)$ is
the hypergeometric function.

For this distribution function we computed the maximum growth rates of the
upper-hybrid waves in dependence on the ratio
$\omega_\mathrm{p}/\omega_\mathrm{B}$ for $\theta_\mathrm{c} = 50^\circ$,
$\kappa=1.5$, $s=16$ and $s=15$,  and $p_\kappa = p_\mathrm{m}$ corresponding to 0.3 $\mathrm{c}$.

\begin{figure}
    \centering
    \includegraphics[width=8cm]{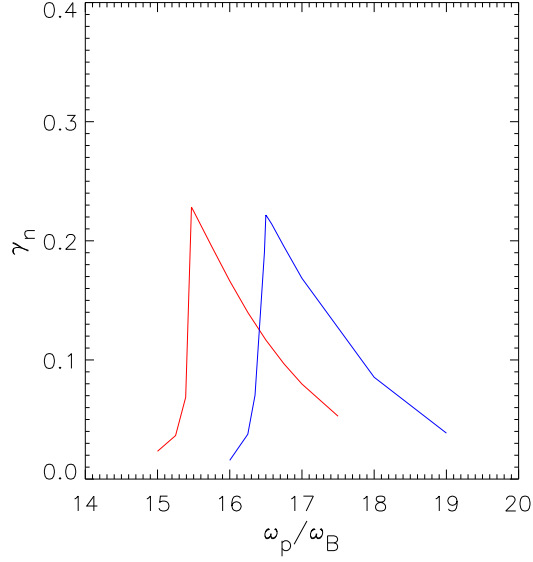}
    \caption{Maximum growth rate of the upper-hybrid waves for the kappa momentum distribution (14) in dependence
    on the $\omega_\mathrm{p}/\omega_\mathrm{B}$. The kappa index is $\kappa$ = 1.5, the gyro-harmonic number $s$ = 15 (red line)
    and $s$ = 16 (blue line),    the loss-cone is $\theta_\mathrm{c}$ = 50$^\circ$, $p_\kappa=p_\mathrm{m}$ corresponds to 0.3 c.}
    \label{figure7}
\end{figure}

Figure~\ref{figure7} shows that the kappa distribution, bounded at small
momentum values, also yields peaks in the spectrum of the growth rate which are
distinctly isolated similarly as the peaks for the power-law distribution
(compare with Figure~\ref{figure2} b).

\section{Frequency Spectrum of Growth Rates}

Zebras are observed in the spectrum in dependence on the radio
frequency, not in the spectrum depending on the ratio
$\omega_\mathrm{p}/\omega_\mathrm{B} $. Therefore, it is of interest to compute
such a frequency spectrum. For it we need to take the plasma frequency from
some observed zebra stripes. In the following computations, we take them for
three stripes of the zebra ($s$ = 25, 26 and 27) observed in the 1 August flare
~\citep{2016SoPh..291.2037Y}: $f_\mathrm{p}$ = $1.344\times10^9$ Hz for $s$ =
25, $f_\mathrm{p}$ = $1.323\times10^9$ Hz for $s$ = 26 and $f_\mathrm{p}$ =
$1.301\times10^9$ Hz for $s$ = 27. To obtain the frequency spectrum, it is
necessary to integrate the above derived growth rates with respect to
$\omega_\mathrm{p}/\omega_\mathrm{B}$, setting for each band its own plasma
frequency

\begin{equation}
\bar{\gamma}_\mathrm{n}=\int \gamma_\mathrm{n}\, \mathrm{d} \left(\frac{\omega_\mathrm{p}}{\omega_\mathrm{B}} \right).
\end{equation}

\begin{figure}
    \centering
    \includegraphics[width=8cm]{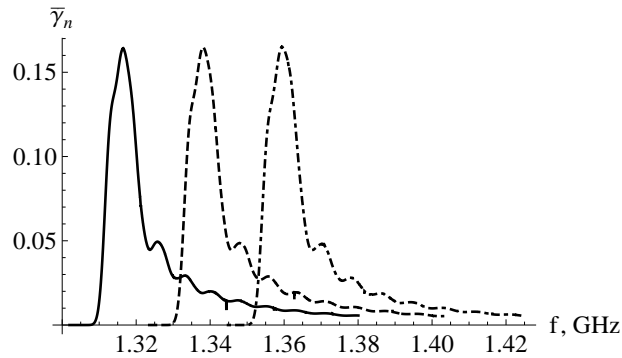}
    \caption{Dependence of the growth rate of the upper-hybrid waves
    on the frequency $f$ for the distribution (14),  with $\kappa$ = 1.5,  $\theta_\mathrm{c}=50^\circ$, $p_\mathrm{m}$
    corresponds to the velocity 0.3 c. The solid line is for $s$ = 25, the dashed line for $s$ = 26 and the dot-dashed line for $s$ = 27.}
    \label{f8}
\end{figure}

Figure~\ref{f8} shows the dependence of the growth rate of the upper-hybrid
waves on the frequency $f$ for the distribution (14),  with $\kappa$ = 1.5,
$\theta_\mathrm{c}=50^\circ$, $p_\mathrm{m}$ corresponds to the velocity 0.3 $\mathrm{c}$. Here and
in the following, the value of $p_\mathrm{\kappa}$ is taken as corresponding to
the temperature of the thermal plasma ($T_\mathrm{e}$=3 $\times$ 10$^6$ K). As seen here
the spectrum shows distinct isolated peaks giving in the spectrum distinct
stripes. An analogous result was also obtained for the power distribution, not
shown here.

\begin{figure}
    \centering
    \includegraphics[width=8cm]{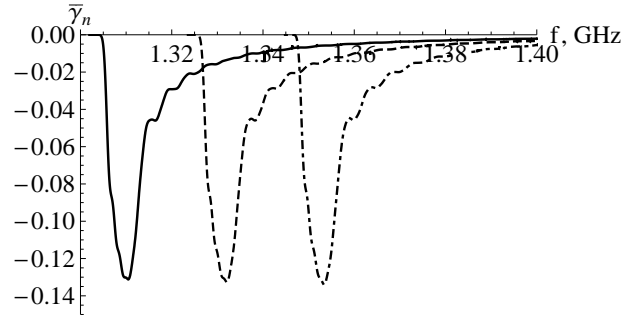}
    \caption{Dependence of the growth rate of the upper-hybrid waves
    on the frequency $f$ for the distribution (14),  with $\kappa$ = 1.5,  $\theta_\mathrm{c}=80^\circ$, $p_\mathrm{m}$
    corresponds to the velocity 0.3 c. The solid line is for $s$ = 25, the dashed line for $s$ = 26 and the dot-dashed line for $s$ = 27.}
    \label{f9}
\end{figure}

Now, let us now check if the conclusion about the significant influence
of the value of the pitch-angle boundary on the frequency spectra is valid.
Figure~\ref{f9} shows the frequency spectrum of the growth rate of the upper-
hybrid waves for the distributions (14) with $\theta_\mathrm{c} = 80^\circ$, $\kappa$ = 1.5
and $p_\mathrm{m}$ corresponding to velocity 0.3 $\mathrm{c}$. As seen here, similarly
as in the case with the power-law distribution, the growth rates are negative.
Note that the negative spectral peaks are well separated from each other.

\begin{figure}
\centering
\includegraphics[width=8cm]{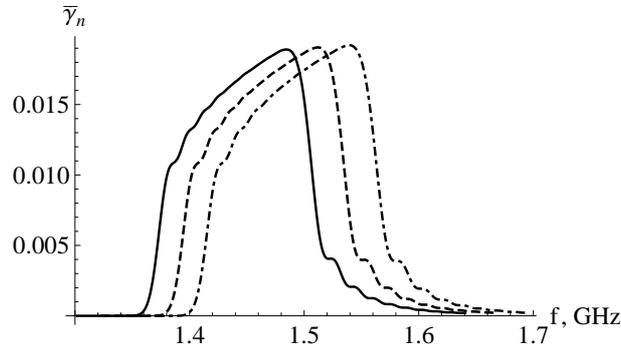}
\caption{Dependence of the growth rate of the upper-hybrid waves
    on the frequency $f$ for the distribution (14),  with $\kappa$ = 1.5,  $\theta_\mathrm{c}=10^\circ$, $p_\mathrm{m}$
    corresponds to the velocity 0.3 c. The solid line is for $s$ = 25, the dashed line for $s$ = 26 and the dot-dashed line for $s$ = 27.}
\label{f10}
\end{figure}

Figure~\ref{f10} shows the frequency spectrum of the growth rate of the
upper-hybrid waves for the distribution (14) with $\theta_\mathrm{c} =
10^\circ$, $\kappa$ = 1.5 and $p_\mathrm{m}$  corresponding to velocity 0.3 c.
Now the spectral bands are very broad. As a result, these bands merge and thus
not forming isolated peaks which are necessary for zebra stripes generation.

\begin{table}
\caption{Frequencies of the growth rate maxima and their frequency differences
in dependance on $\theta_\mathrm{c}$ and $v_\mathrm{m}$ for the gyro-harmonic
numbers $s$ = 25, 26 and 27.} \centering
\begin{tabular}{ccccccc}
\hline
\hline
 $\theta_\mathrm{c}$ & $v_\mathrm{m}$ & $f_{max}^{s=25}$  & $f_{max}^{s=26}$ & $f_{max}^{s=27}$ & $\Delta f (s=25-26)$ & $\Delta f (s=26-27)$  \\
      ($^\circ$)     &                &    (GHz)          &    (GHz)         &    (GHz)         &      (GHz)           &      (GHz) \\
\hline
50 & 0.2 $c$ & 1.381 & 1.359 & 1.337 & 0.022 & 0.022 \\
50 & 0.3 $c$ & 1.359 & 1.338 & 1.317 & 0.021 & 0.021 \\
50 & 0.4 $c$ & 1.365 & 1.347 & 1.328 & 0.018 & 0.019  \\
30 & 0.3 $c$ & 1.379 & 1.358 & 1.336 & 0.021 & 0.022 \\
65 & 0.3 $c$ & 1.356 & 1.334 & 1.313 & 0.022 & 0.021 \\
\hline \label{tab2}
\end{tabular}
\end{table}

Finally, we computed the growth rate of the upper-hybrid waves for the
distribution (\ref{ee2}) in dependance on frequency with  $\kappa$ = 1.5, for
different $\theta_\mathrm{c}$ and $p_\mathrm{m}$ (corresponding to the velocity
$v_\mathrm{m}$). Other values in computations were taken the same as for
Figure~\ref{f8}. In Table~\ref{tab2} we show the frequencies of the growth rate
maxima and frequency differences of their maxima. It looks that for
$\theta_\mathrm{c} = 50^\circ$ the frequency difference slightly decreases with
the increase of $v_\mathrm{m}$. For fixed $v_\mathrm{m}$ this difference
(within errors of computations corresponding to the last number in~$\Delta f$)
practically does not depends on $\theta_\mathrm{c}$.

\section{Discussion and conclusions}
It was shown that the growth rate of the upper-hybrid waves for the power-law
momentum distribution with the low-momentum cut-off and the loss-cone
anisotropy strongly depends on the pitch-angle boundary. The maximum growth
rate was found for the pitch-angle $\theta_\mathrm{c} \approx 50^\circ$. For
small angles the growth rate is broad and flat and for high pitch-angles even
the absorption occurs.

We made computations for two power-law indices for $\delta$ = 5 and 10. While
for the power-law index $\delta$ = 5 the zebra structure can be formed only in
limited interval of $v_\mathrm{m}$ around the velocity 10 $v_\mathrm{T}$ = 6.75
$\times$ 10$^7$ m s$^{-1}$, in the case with $\delta = 10$ the width of the
growth rates are two time smaller and thus more favorable for the zebra pattern
generation. Positions of the growth rate maxima in both the cases are
approximately the same. These results agree to those of~\cite{2007SoPh..241..127K}.

An analysis of the growth rate of the upper-hybrid waves for the anisotropic
kappa momentum distribution for all electron momenta ($p > 0$) (which is a~contribution to the dense background plasma) shows: a) When we decrease the
characteristic momentum $p_\kappa$ then the maximum of the growth rate is
shifted to lower values of $\omega_\mathrm{p}/\omega_\mathrm{B}$. b) The growth
rates for the kappa-distribution with $\kappa = 1.5$ and $\kappa = \infty$
shows a similar behavior, but values of the growth rates for $\kappa =
\infty$ are a little bit higher. It is due to that in this case the
distribution with $\kappa = \infty$ has more electrons for low momentum
values than that with $\kappa = 1.5$. c)~The frequency widths of the growth
rate maxima are very broad. We also found that the frequency difference
between the frequencies of the growth rate maxima slightly decreases with the
increase of $v_\mathrm{m}$. For fixed $v_\mathrm{m}$ this difference
practically does not depends on $\theta_\mathrm{c}$.

But, if we take a more realistic distribution, namely, the single kappa
distribution which is isotropic up to some large momentum $p_\mathrm{m}$ and anisotropic
with loss-cone above this momentum then distinct peaks of the growth rate
appear and thus distinct zebra stripes can be generated. It means that the
restriction for small momenta for the anisotropic part of distributions
(power-law or kappa) is of principal importance for the zebra stripes
generation.

For the first time, the dependence of the growth rate on the radio frequency
was computed. In this case the spectral peaks are much more distinct than in~the case of the dependence of the growth rate on the ratio of the plasma and
cyclotron frequencies. Thus, analyzing observed radio spectra, we can assume
smaller values of the power-law or kappa indices.

Note that for high values of the pitch angle anisotropy, where the
absorption occurs, the inverse zebra stripes can be produced on some radio
continua.


\begin{acks}
We thank an anonymous referee for valuable comments. M. Karlick\'y
acknowledges support from Grants 17-16447S and 18-09072S of the Grant Agency of
the Czech Republic. L.V. Yasnov acknowledge support from Grant 18-29-21016-mk
and partly from Grant 18-02-00045 of the Russian Foundation for Basic Research.
This work was supported by The Ministry of Education, Youth and Sports from the
Large Infrastructures for Research, Experimental Development and Innovations
project IT4Innovations National Supercomputing Center LM2015070.
\end{acks}

%
\bibliographystyle{spr-mp-sola}
\bibliography{L2018}
\end{article}
\end{document}